# PEPX-type lattice design and optimization for the High Energy Photon Source

JIAO Yi (焦毅), XU Gang（徐刚）

Institute of High Energy Physics, CAS, Beijing 100049, P.R. China

**Abstract:** A new generation of storage ring-based light source, called diffraction-limited storage ring (DLSR), with the emittance approaching the diffraction limit for multi-keV photons by using the multi-bend achromat lattice, has attracted worldwide and extensive studies of several laboratories, and been seriously considered as a goal of upgrading the existing facilities in the imminent future. Among various DLSR proposals, the PEPX design [*Cai Y, et al. Phys. Rev. ST Accel. Beams, 2012, 15: 054002*] based on the 'third-order achromat' concept and with the special design of a high-beta injection straight section demonstrated that, it is feasible to achieve sufficient ring acceptance for off-axis injection in a DLSR. For the High Energy Photon Source planned to be built in Beijing, PEPX-type lattice has been designed and continuously improved. In this paper, we report the evolution of the PEPX-type design, and discuss the main issues relevant to the linear optics design and nonlinear optimization.

## 1 Introduction

Along with the progress in accelerator technology and the growing requirements of brighter photon flux, a new generation of storage ring-based light source, named diffraction-limit storage ring (DLSR, sometimes referring to as 'ultimate storage ring') have been extensively designed and studied (see Ref. [1] for an international overview). By pushing down the emittance to approach the diffraction-limit for the range of x-ray wavelengths of interest for scientific community (e.g. ~100 pm.rad for λ = 1 nm and ~10 pm.rad for λ = 0.1 nm, with λ being the X-ray wavelength), DLSR can provide significantly higher coherent photon flux than presently available.

The horizontal natural emittance $\varepsilon_0$ can be expressed as [2]

$$\varepsilon_0 = C_q \gamma^2 \frac{F(type)}{J_x} \theta^3, \tag{13}$$

where $C_q = 3.83 \times 10^{-13}$ m, $\gamma_L$ is the Lorenz factor, $J_x$ is the horizontal damping partition number, the value of $F(type)$ depends on the lattice type [e.g., the minimum $F$ factor for a double-bend achromat (DBA) and a seven-bend achromat (7BA) are ~0.065 and ~0.034, respectively], and $\theta$ is the bending angle of the dipole.

Evidently, to reduce the emittance to a very small value, e.g., several tens of pm.rads, the most effective way is to decrease the bending angle of the dipole, or namely, to increase the number of the dipole $N_d$, which, however, implies a large circumference (∝ ~ $N_d$) and hence a high budget. To control the budget to a reasonable level and meanwhile, to minimize the emittance as much as possible, multi-bend achromats with compact layout are usually adopted in DLSR designs. On the other hand, to achieve ultralow emittance, it requires strong transverse focusing and strong chromatic sextupoles as well, bringing about significant nonlinearities which, if not well controlled, may leads to a poor performance of the light source. Thus, a delicate optimization of the nonlinear terms is necessary to obtain sufficient dynamic aperture (DA) and momentum acceptance (MA) for an efficient injection and a long enough Touschek lifetime. It has been emphasized since one decade ago [3] that nonlinear optimization and linear matching are coupled, and phase optimization helps to minimize the nonlinear effects. For a DLSR, phase optimization indeed becomes an indispensable tool for achieving a satisfy ring performance. In the PEPX design based on a 7BA lattice [4] with $E_0 = 5$ GeV ($E_0$ is the nominal beam energy), $\varepsilon_0 = 30$ pm.rad, the phase advance of each 7BA is chosen in such a value that every eight 7BA forms a third-order achromat, where the nonlinear driving terms up to 4th-order can be largely cancelled; in addition, a special section is designed with large $\beta_x = 200$ m (while keeping the phase advance the same as that of a normal section) to ensure large enough acceptance (~ 10 mm in $x$ plane) for off-axis injection. This design philosophy, in the past few years, have inspired studies of designing lattice in a similar scheme for the High Energy Photon Source (HEPS, originally referring to as BAPS), a kilometer-scale, 5-6 GeV light source that is planned to be built in Beijing. In the following, we will review the evolution of the PEPX-type lattice design for HEPS in Sec. II, and present the optimized DA and MA for the latest version in Sec. III. For this type of design, it is noted that the specific design of the injection section causes intrinsic sensitivity of the ring optics to the momentum deviation (hereafter denoted by δ), and hence increases the difficulty in pursuing a large enough MA. Even so, we show that this problem can be overcome to a large extent through careful tuning of the strengths of multi-families of sextupoles and octupoles. Conclusive remarks are given in Sec. IV.

## 2 PEPX-type lattice designs for HEPS

As early as 2010, a baseline design of the HEPS with 48 DBAs was proposed to provide 5-GeV electron beam with $\varepsilon_0$ of 1.6 nm.rad, and of 0.5 nm by using additional damping wigglers, within a circumference of $C = 1209$ m [5]. Since then, efforts have been made to look for alternative lattice designs with much lower emittance.

The first attempt [6] was to reduce the horizontal natural emittance to 75 pm.rad by use of a PEPX-type lattice consisting of thirty two 7BAs. To control the circumference and hence the total cost, several measures were adopted, such as using modified-TME unit cells with horizontally defocusing gradient combined in the dipole (resulting in $J_x > 1$ for even lower emittance) [7], and utilizing small-aperture magnets and vacuum systems (with magnet bore radii of 12.5 mm, following MAX-IV design [8]) enabling multipole gradients of up to 47 T/m and 7700 T/m$^2$ (for the pole face field of 0.6 Tesla). Finally a compact layout was reached with each modified TME-unit cell of 3.8 m and with $C = 1263.4$ m. In this lattice, high-beta and low-beta straight sections are alternatively distributed, for the sake of efficient injection and high-coherence flux emission from insertion device (ID), respectively. Similarly to the PEPX design, the linear optics was matched such that every sixteen 7BAs forms a quasi-3rd-order achromat. This greatly facilitates the subsequent nonlinear optimization, where a theoretical analyzer based on the perturbation theory was used to minimize the nonlinear terms. Finally, it was feasible to achieve a moderate integer resonance-clear acceptance (~ 6 mm in $x$ plane and ~ 2 mm in $y$ plane) for off-axis injection and a large MA of $\delta_m \sim 3\%$. More details can be seen in Ref. [6].

However, it was noted that in this design only half of the straight sections were designed with low beta functions for optimal matching of the electron and photon beam ($\beta \sim L_{ID}/\pi$ [9] or $L_{ID}/2\pi$ [4], with $L_{ID}$ being the ID length). To provide as many ID sections with optimal beta functions as possible, a modified PEPX-type lattice composed of thirty six 7BAs with $\varepsilon_0$ = 51 pm.rad and $C$ = 1364.8 m was proposed [10]. In this modified design, the length of each modified TME-unit cell was further reduced to 3.6 m; every twelve 7BAs forms a quasi-3rd-order achromat; and most importantly, among 36 straight sections, 34 of them were designed specifically for ID (see Fig. 1) while the other two were modified to have large beta functions and longer drift space (9.6 m vs. 7 m) to accommodate injection devices and RF cavities (see Fig. 2). To restore the periodicity, the phase advance of the high-beta section was tuned to be same as that of a normal section, or with a difference of $2n\pi$ ($n$ is integer). After nonlinear optimization, we were able to achieve a moderate integer resonance-clear acceptance (~ 7 mm in $x$ plane and ~ 5 mm in $y$ plane), however, with a relatively small MA ($\delta_m$ ~ 1.5%). This is because that the difference in phase advance between two kinds of sections will deviate from $2n\pi$ as $\delta$ increases, and the periodicity will be destroyed. In such a case, resonances will be more excited, leading to unstable motion for the particles with large momentum oscillations.

Most recently, the HEPS lattice is required to be designed with a fixed circumference of $C$ = 1296 m (with +/−3 m varying range) to obtain a harmonic number of 2160 (for ~500 MHz RF cavities), and to reserve the feasibility of ramping the beam energy from 5 to 6 GeV in a future upgrade. In addition, it was noticed that the available multipole gradient can be further enhanced (e.g. ~100 T/m for quadrupole [11]), if using high-permeability pole material (e.g. vanadium permendur) or permanent magnet material near the poles to reduced saturation.

Based on above, a latest version of PEPX-type lattice consisting of forty four 7BAs was designed, with $\varepsilon_0$ = 88 (or 61) pm.rad for 6 (or 5) GeV and $C$ = 1294.2 m. In this design, due to adoption of high-gradient quadrupole/sextupoles, more compact layout is reached than previously, with each modified TME-unit cell of 3 m (see Fig. 3). Forty 7BAs are in standard design; the phase advance of each is chosen to be $\mu_x = 4\pi + \pi/4 + \delta\upsilon_x*\pi/20$ and $\mu_y = 2\pi + \pi/4 + \delta\upsilon_y*\pi/20$ (with $\delta\upsilon_x$ and $\delta\upsilon_y$ being the decimal portions of the nominal working point), such that every eight 7BAs constitute a quasi-3rd-order achromat. Slightly different from the second version, the other four 7BAs are designed to provide 10-m long straight sections, with no sextupole/octupole and with phase advance of $2n\pi$, which yields an identity linear transformation and hence restore the periodicity. This special design has similar drawback to the second version, i.e., with great difficulty in enlarging the MA of the ring. Nevertheless, by using multi-families of harmonic sextupole and octupoles and with careful optimization, large DA or large MA can be obtained with two separate modes, which are only different in sextupole/octupole strengths and hence can be easily switched from one to another (details will be presented in the next section).

Besides, it should be mentioned that associated with the decreasing dispersions and increasing natural chromaticities as reducing the emittance to even smaller emittance, PEPX-type lattice design may approach its ultimate performance from a practical standpoint, otherwise extremely high-gradient or rather thick sextupoles will be required. This problem can be greatly alleviated by use of a hybrid-MBA lattice where the chromatic sextupoles are located in a sector with a high-dispersion bump, as described in Ref. [11]. HEPS lattice Design of this type is actually synchronized with the PEPX-type design, which, however, will be introduced elsewhere.

To aid the reader, and to illustrate the evolution of the PEPX-type lattice design for HEPS in an intuitive manner, the main parameters of the three versions of lattices are listed in Table 1.

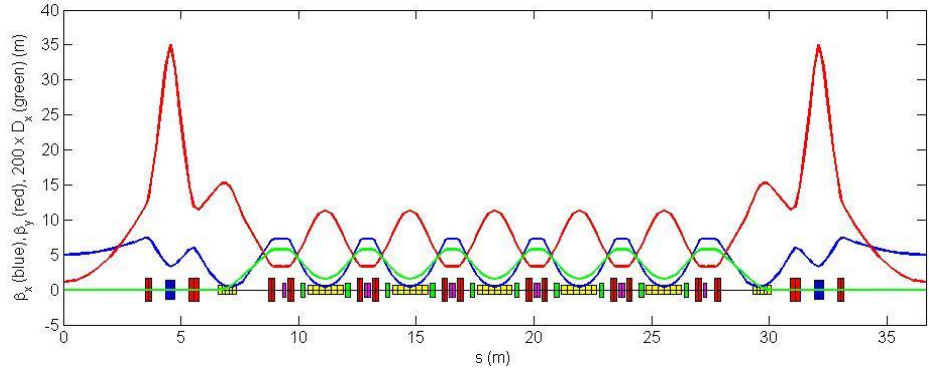

Fig. 1. (color online) Layout and optical functions of a standard 7BA with 7-m ID straight section, for the second version of the PEPX-type lattice for HEPS.

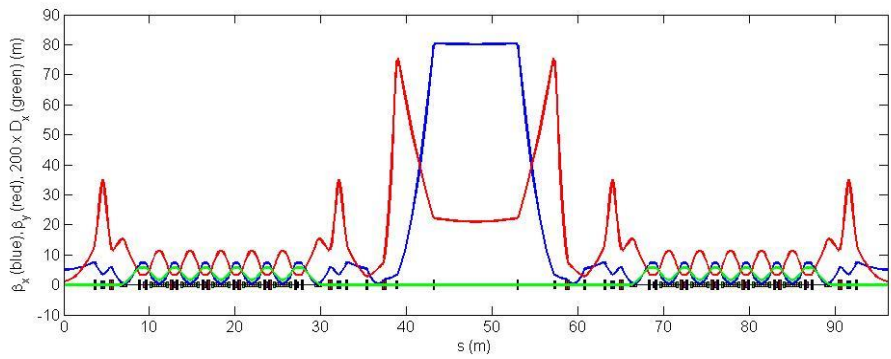

Fig. 2. (color online) Layout and optical functions of two 7BAs with 9.6-m long straight section in between, for the second version of the PEPX-type lattice for HEPS. The optics is matched in such a way that the phase advance of the 9.6-m straight section is different from that of the normal 7-m straight section by $\Delta\mu_x = 2\pi$ and $\Delta\mu_y = 0$.

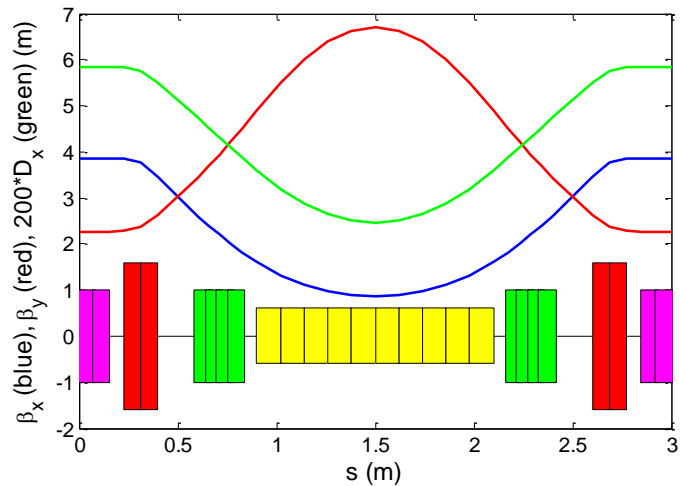

Fig. 3. (color online) Layout and optical function of the 3-m unit cell in middle of the standard 7BA, for the latest version of the PEPX-type lattice for HEPS. Purple blocks represent horizontally focusing sextupoles, red bocks represent horizontally focusing quadrupole, green blocks represent horizontally defocusing sextupoles, and yellow block represents dipole combined with horizontally defocusing gradient.

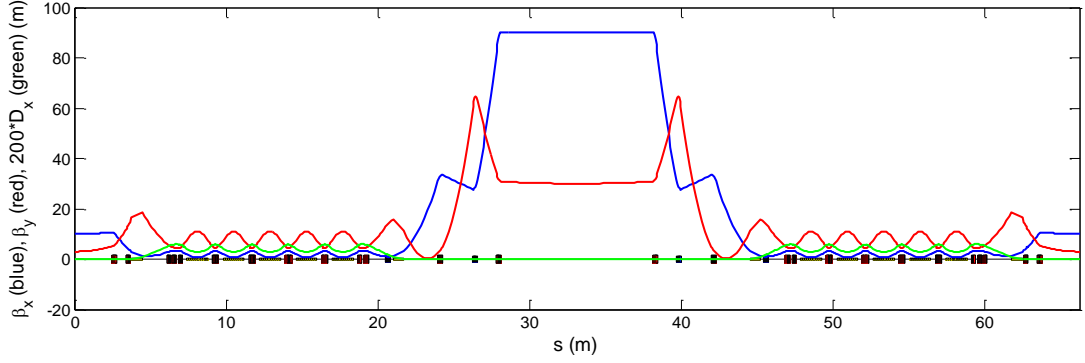

Fig. 4. (color online) Layout and optical functions of two specially designed 7BAs with 10-m long straight section in between, for the latest version of the PEPX-type lattice for HEPS. The phase advance of these two 7BAs is set to $\mu_x = 4\pi$ and $\mu_y = 2\pi$.

Table 1. Main parameters of the three version of the PEPX-type lattice for HEPS

| Parameters | The first version | The second version | The third version |
|---|---|---|---|
| Energy $E_0$ (GeV) | 5 | 5 | 6 (5) |
| Circumference $C$ (m) | 1263.4 | 1364.8 | 1294.2 |
| Horizontal damping partition number $J_x$ | 1.40 | 1.54 | 2.05 |
| Natural emittance $\varepsilon_0$ (pm.rad) | 75 | 51 | 88 (61) |
| Number of 7BA achromats | 32 | 36 | 44 |
| Maximum quadrupole gradient (T/m) | 48 | 48 | 80 |
| Maximum sextupole gradient (T/m$^2$) | 7700 | 7700 | 11000 |
| Number/length (m) of high-beta straight sections | 16/10 | 2/9.6 | 2/10 |
| Beta functions (m) in high-beta straight section (H/V) | 41/4.7 | 80/21 | 90/30 |
| Number/length (m) of low-beta straight sections | 16/6 | 34/7 | 42/5 |
| Beta functions (m) in low-beta straight section (H/V) | 4.5/1.7 | 5/1.11 | 10/3 |
| Working point (H/V) | 98.40/34.30 | 113.39/39.30 | 93.14/49.40 |
| Natural chromaticities (H/V) | -189/-113 | -184/-181 | -112/-107 |
| Damping times (ms, x/y/z) | 20/28/17.4 | 27.5/42.4/29.1 | 11.2/23.0/24.3 |
| Energy spread $\sigma_\delta$ | $8 \times 10^{-4}$ | $7 \times 10^{-4}$ | $1 \times 10^{-3}$ |
| Momentum compaction $\alpha_p$ | $3.86 \times 10^{-5}$ | $4.0 \times 10^{-5}$ | $5.9 \times 10^{-5}$ |

## 3 Nonlinear optimization of latest design

It is thought that the DA (for on-momentum particles) is most relevant to the capture of the injected beam into the ring. In the case of off-axis injection, the DA (in the plane of the injection) should be large enough (e.g., >~ 5 mm) to allow a high capture efficiency. MA is most relevant to the Touschek lifetime, which is the main limitation of the available beam lifetime in modern light sources, especially in a DLSR. For PEPX design, an empirical relation between the Touschek lifetime and the MA is found to be $T \propto \delta_m^5$ ($T \sim 2.5$ hours with $\delta_m = 2\%$, and $T$ will increase to ~21 hours if with $\delta_m = 3\%$ [4]). Optimizing the beam dynamics to obtain sufficient DA and MA is of critical importance for a DLSR.

In our study, the particle motion in presence of the nonlinearities originating from chromatic sextupoles is studied in an analytical manner. By using a program based on perturbation theory, we can derive the nonlinear terms [such as the detunning (up to the 2nd order), high-order chromatic (up to the 4th order), and resonance driving terms (up to the 6th order)] as polynomial functions of the sextupole/octupole strengths. With these polynomial functions one can quickly evaluate the nonlinear terms for different set of sextupole/octupole strengths, which makes it feasible to perform MOGA (multi-objective genetic algorithm [12]) optimization with a large range scan in multi-variable space, an most importantly, in a reasonable time (a few hours on a single PC). In this way, we can obtain the so-called Pareto optimal front that includes multi-solutions (instead of one) showing all the possible tradeoffs between the different objectives. Note that the perturbation theory is based on the assumption of small particle offset and momentum deviation from the reference particle. The solutions from the analytical approach may fail to perform well for large transverse amplitudes and momentum deviation (especially for a DLSR with extremely strong nonlinearities). Therefore it is necessary to verify the realistic performances (DA and MA) of the lattice with these results by means of numerical tracking and frequency map analysis [13]. Usually iterations between tacking and analytical optimization is required to achieve large enough DA and MA at last.

As mentioned, in the latest design, sextupoles and octupoles are located only in the standard 7BAs, where the sextupoles are grouped in eleven families (with 2 families of harmonic ones) and the octupoles are grouped in three families. With iterations of the above optimization routine, one solution promising large horizontal DA (larger than the physical aperture of 11 mm) and robust dynamics is finally obtained. The corresponding DA and FM are shown in Fig. 5. It is found that the half integer resonance $2\upsilon_x = 187$ and $\upsilon_x = 94$ are reached at a large $x$ amplitude, i.e., −11.4 mm and −13 mm (or at even larger positive amplitudes); for the particle motion with $|x| < 11$ mm, no dangerous low-order resonance is crossed, and only a few high-order resonances, $4\upsilon_x + 4\upsilon_y = 570$ and $7\upsilon_x = 652$, have rather weak impact on the dynamics. In addition, the tune shifts with amplitude and with δ calculated from numerical tracking and from theoretical analysis are plotted in Fig. 6. One can see that the analyzer succeeds to predict the tune shifts with amplitude for on-momentum particles, but fails to predict the tune shifts with δ at relatively large momentum deviation (δ ~ 1%). This is because that the nonlinear terms are derived by assuming the identity transformation of the injection section still holds as the momentum deviation increases, which does not reflect the actual circumstances. It appears that this mode (denoted by mode 1) can be used for injection, while for storing the beam more efforts are needed to enlarge the MA.

To this end, we use directly the tune shifts with δ calculated numerically (this can be done in a very short time) as the optimization goal, and search for the result promising a much less tune shift with δ and a fairly good on-momentum dynamics. Finally, we obtain another solution that enables MA of $\delta_m$ = 3%, with, certainly, a price of smaller DA. The tune shifts with δ, and the momentum dependent DA from numerical tracking are presented in Fig. 7, and the on-momentum DA and the corresponding FM are shown in Fig. 8. It shows that for this mode (denoted by mode 2) the coupling resonance $\upsilon_x - \upsilon_y = 44$ and the integer resonance $\upsilon_x = 93$ dominate the beam dynamics in $x$ and $y$ planes, resulting in a smaller resonance-clear acceptance (~ 7 mm in $x$ plane and ~ 5 mm in $y$ plane). Such an acceptance may be not sufficient for off-axis injection, but is still much larger than the equilibrium transverse size of the beam after injection (with $\sigma_x$ ~ 80 μm and similar or even smaller $\sigma_y$).

Further optimizations are performed and it seems scarcely possible to obtain a mode with both large DA (with similar size to mode 1) and large MA (with similar $\delta_m$ to mode 2). Nevertheless, one can use mode 1 during injection, and then switch to mode 2 (it may take a few seconds) for a long enough Touschek lifetime. Since the linear optics remains the same for these two modes, and only sextupole/octupole magnets need to be ramped, it is believed that the dynamics will keep stable during mode switching. Actually, this has been verified with numerical tracking, with the result shown in Fig. 9.

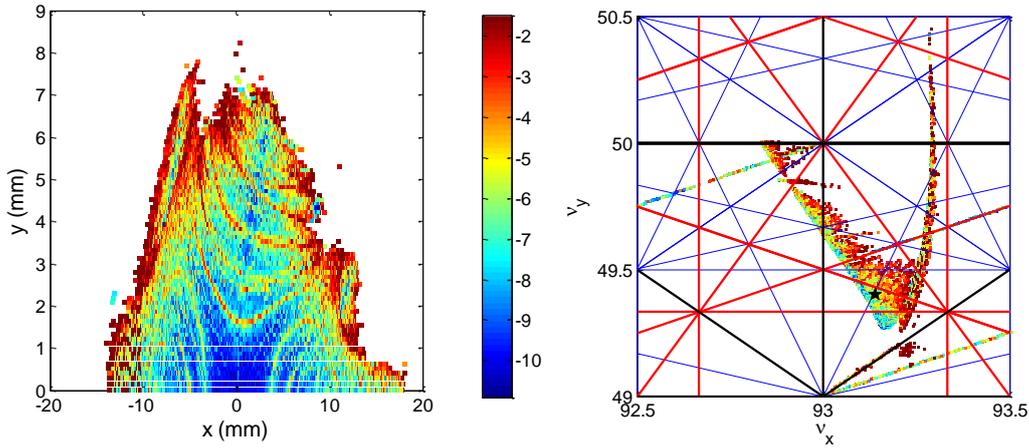

Fig. 5. (color online) The dynamic aperture and frequency map obtained after over 1024 turns for latest version of the PEPX-type lattice for HEPS (mode 1). The colors, from blue to red, represent the stabilities of the particle motion, from stable to unstable.

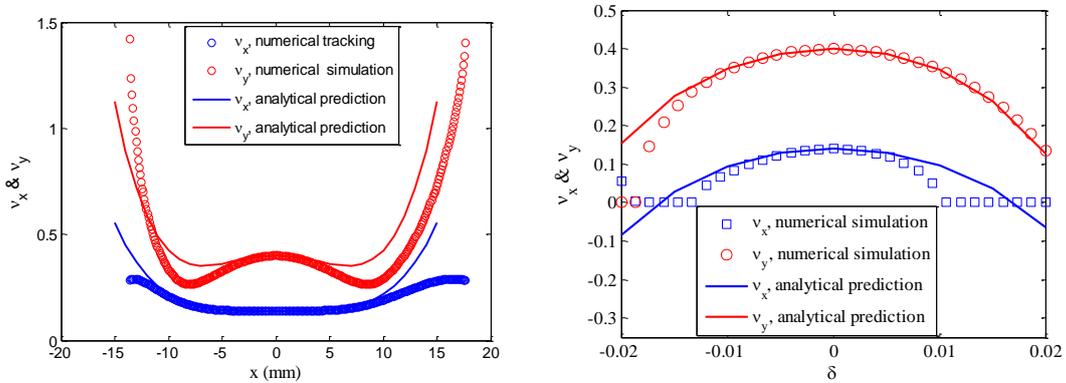

Fig. 6. (color online) Tune shifts with horizontal amplitude (left) and with momentum deviation (right), extracted from the numerical tracking results and the theoretical results, respectively, for latest version of the PEPX-type lattice for HEPS (mode 1).

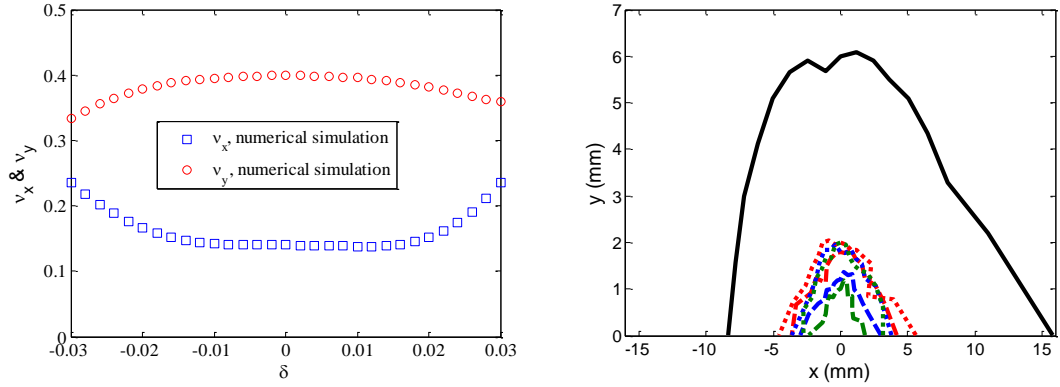

Fig. 7. (color online) Simulation results of tune shifts with momentum deviation (left) and of the momentum dependent DA, for latest version of the PEPX-type lattice for HEPS (mode 2). In the right plot, with $\delta = 0$ (black solid), $\delta = 1\%$ (red dashed), $\delta = -1\%$ (red dotted), $\delta = 2\%$ (blue dashed), $\delta = -2\%$ (blue dotted), $\delta = 3\%$ (green dashed) and $\delta = -3\%$ (green dotted).

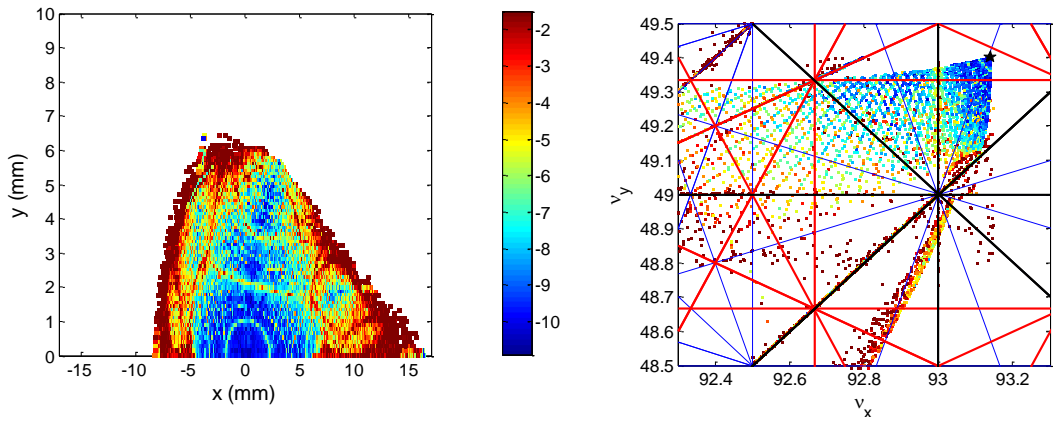

Fig. 8. (color online) The dynamic aperture and frequency map obtained after over 1024 turns for latest version of the PEPX-type lattice for HEPS (mode 2, with sextupole/octupole strengths different from those for Fig. 5). The colors, from blue to red, represent the stabilities of the particle motion, from stable to unstable.

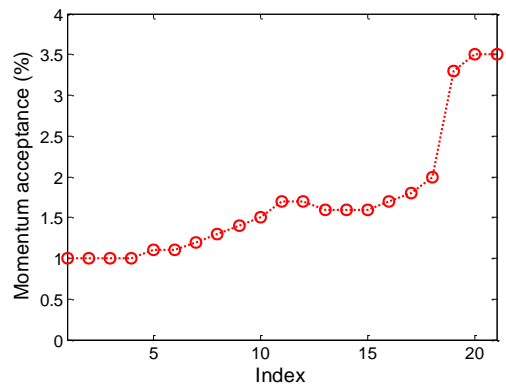

Fig. 9. (color online) Simulation results of the MA variation during the mode switching (assuming the magnets are ramped in 21 steps), for latest version of the PEPX-type lattice for HEPS.

## 4 Conclusions

In this paper we show the PEPX-type lattice designs for HEPS composed of quasi-3rd-order achromats. To provide as many as possible ID sections with a fixed circumference (~1300 m) and with minimized emittance (less than 100 pm.rad), small-aperture, high-gradient multipoles and most importantly, a specal high-beta injection section (with phase advance or difference in phase advance of $2n\pi$) are adopted in the design. Such a design helps to restore the periodicity and turns to be important for achieving a ring acceptance of about 10 mm in horizontal plane for off-axis injection. The disadvantage is also obvious that the periodicity will be destroyed for the off-momentum particles, leading to intrinsic difficulty in pursuing a large enough momentum acceptance, which, however, can be overcome by delicate nonlinear optimization.

## Acknowledgement

This work is supported by the National Natural Science Foundation of China (No. U1332108).